\documentstyle[prl,aps,preprint,tighten,floats,aps,epsf,psfig]{revtex}

\def\abstract{\if@twocolumn
\section*{Abstract}
\else \normalsize
\begin{center}
{\bf Abstract\vspace{0pt}}
\end{center}
\setlength{\baselineskip}{4ex}
\quotation
\fi}
\def\endabstract{\if@twocolumn\else\endquotation\fi}

\begin{document}
\draft
\preprint{UPR-0780-T, hep-ph/980xxxx}
\date{\today}
\setcounter{footnote}{1}
\title{Neutrino Anti-Neutrino Transitions}
\author{Paul Langacker and Jing Wang}
\address{
          Department of Physics and Astronomy\\
          University of Pennsylvania\\
          Philadelphia, PA  19104\\
}

\begin{titlepage}
\maketitle
\def\thepage {}        % Kill page numbering for title page

\begin{abstract}

We consider transitions between neutrinos and anti-neutrinos in laboratory experiments in five scenarios. These include the case in which the helicity flips, producing an anti-neutrino with normal weak interactions, and helicity preserving oscillations into an $SU(2)$ singlet state which only interacts by mixing or new interactions. The ratio of $\mu^{+}$ and $\mu^{-}$ events for a high energy $\nu_{\mu}$ beam from pion decay rescattered from a nucleon target, and the ratio between $e^{+}$ and $e^{-}$ events for a rescattered low energy $\nu_{e}$ beam are calculated in each case. The upper limit on the ratio is about $10^{-7}\sim 10^{-12}$ for a high energy $\nu_{\mu}$ beam and $10^{-4} \sim 10^{-14}$ for a low energy $\nu_{e}$ beam, too small to observe in present experiments.  

\end{abstract}
\end{titlepage}

\section{Introduction}

Neutrino anti-neutrino transitions have been predicted by various non-standard models \cite{s1}$-$\cite{s5A}. Such transitions for real neutrinos are complementary to the lepton-number-violating transitions of virtual neutrinos that are searched for in neutrinoless double beta decays \cite{s5B}. The latter are extremely sensitive to $\nu_{e}$ effects, but blind to other lepton flavors. Recently, interest in real transitions revived due to the discussion of possible astrophysical applications of spin precession in a transverse $B$ field for Majorana neutrinos \cite{s6}. In this work, we study possible scenarios of neutrino anti-neutrino transitions in the context of the standard model and its allowed extensions and consider the possibility of observations in laboratory experiments. 

We consider mechanisms for lepton number violation that involve a spontaneously generated Majorana neutrino mass. Several processes can lead to helicity flip, in which left-handed neutrinos, $\nu_{L}$, are converted into right-handed anti-neutrinos, $\nu_{R}^{c}$. Neutrinos with Majorana masses can be produced/annihilated in the wrong helicity states by the ordinary left-handed charged weak currents because $\nu_{L}$ and $\nu_{R}^{c}$ are connected by the Majorana mass in the Lagrangian. However, the amplitude is proportional to $(m_{\nu}/E_{\nu})$, and is therefore small. The transition magnetic moments of a Majorana neutrino can connect $\nu_{i L}$ and $\nu_{j R}^{c}$ from different families. Hence, a transverse $B$ field can rotate $\nu_{i L}$ into $\nu_{j R}^{c}$ \cite{s5A},\cite{s6}. However, the transition probability is proportional to $|\mu_{\nu} B|^{2}$, which is small. Models in which lepton number is spontaneously broken by the vacuum expectation value of an additional Higgs field predict helicity flip neutrino decays: $\nu_{2 L}^{c} \rightarrow \nu_{1 R}^{c} + \chi$, where $\chi$ is the Goldstone boson (Majoron) associated with lepton number violation \cite{s7}. However, the decay rate, depending on the Yukawa coupling constant of the extra Higgs field and the mixing angle, is restricted to be small.

Other possibilities are helicity non-flip transitions. They can occur when Dirac and Majorana masses are both present. The mass eigenstates become combinations of weak eigenstates $(\nu_{L}, \nu_{L}^{c}, \nu_{R}, \nu_{R}^{c})$, where $\nu_{L}$ and $\nu_{R}^{c}$ are respectively the left and right chiral components of an ordinary $SU(2)$ doublet (active) neutrino. Similarly, $\nu_{L}^{c}$ and $\nu_{R}$ are the left and right chiral components of an $SU(2)$ singlet (sterile) neutrino. For comparable Dirac and Majorana masses, the mismatch between weak and mass eigenstates results in ($2^{nd}$ class) oscillations \cite{s2}-\cite{s2E} between neutrinos and anti-neutrinos: the probability that an initial $\nu_{L}$ oscillates into $\nu^{c}_{L}$ at time $t$ is \footnote{Pure Dirac or pure Majorana masses lead only to first class oscillations, in which the lepton family is changed (e.g., $\nu_{e L} \rightarrow \nu_{\mu L}$), but neutrinos are not changed into anti-neutrinos.}:
\begin{equation}
P(\nu_{L}\rightarrow \nu^{c}_{L}) = |\langle \nu^{c}_{L}|\nu_{L}(t) \rangle |^{2} = 
\sin^{2}{2\theta} \sin^{2}{\Delta/2} ; 
\end{equation}
where 
 $\Delta=2.54 \frac{(m_{1}^{2}-m_{2}^{2})(eV^{2}) L(m)}{E_{\nu}(MeV)}$, $L$ is the distance traveled, and $\theta$ is the mixing angle. The transition rate can be large for $2^{nd}$ class oscillations. However, oscillations don't change the helicity, and second class oscillations result in a $\nu^{c}_{L}$ which is sterile with respect to standard model interactions. 

Since disappearance experiments have relatively low sensitivity and don't distinguish first and second class oscillations, we consider some non-standard models in which $\nu_{L}^{c}$ can have (suppressed) interactions. In the $SU(2)_{R}\times SU(2)_{L} \times U(1)$ model \cite{s8}, $\nu^{c}_{L}$ and $l^{+}_{L}$ are grouped into an $SU(2)_{R}$ doublet. Hence, $\nu^{c}_{L}$ can be scattered into $l^{+}_{L}$ through the right-handed charged current. However, the effect is suppressed by $M_{W_{R}}^{-2}$ for the right-handed interactions or the small mixing between $W_{L}$ and $W_{R}$. In $SU(2)_{L} \times U(1)$ models with heavy exotic fermions \cite{s10}, the light ordinary neutrinos are mixed with heavy exotic ones. Consequently, the state $\nu_{L}$ associated with $l_{L}^{-}$ in weak decays at low energy is not a complete state and thus not orthogonal to $N_{L}^{c}$, which belongs to an exotic doublet \cite{s10A}. $2^{nd}$ class oscillations can also occur. Both mechanisms lead to lepton number violating processes. However, the effects are strongly suppressed by small mixing between light and heavy fermions.

Our conclusion is fairly negative: the upper limit on the transition probability is probably impossible to observe in present laboratory neutrino experiments, because of the small transition rate for helicity flip cases or the small rescattering rate for helicity non-flip cases.   

The paper is arranged in four sections. In section II, we consider helicity flip transitions; section III discusses non-flip transitions; in section IV, we discuss the results. 

We consider two types of experiments. In the first, a high energy $\nu_{\mu}$ beam from pion decay in flight is deep inelastically rescattered from an isoscalar nucleon target. We calculate the ratio between $\mu^{+}$ and $\mu^{-}$ events after rescattering as a measurement of the transition.\footnote{It is straightforward to generalize to transitions which also change lepton families.} The cross section for deep inelastic neutrino scattering is calculated in the simple parton model. In the second case, we consider a low energy electron neutrino beam rescattered from a nucleon target, and calculate the ratio between $e^{+}$ and $e^{-}$ events after rescattering. For low energy neutrino nucleon scattering, we do not specify the specific nuclear transition, but assume that $r = \sigma^{\bar{\nu}_{e}N} / \sigma^{\nu_{e}N}$ is of order 1, where $\sigma^{\bar{\nu}_{e}N}$/$\sigma^{\nu_{e}N}$ is the cross section of anti-neutrino/neutrino scattering through normal weak interactions. \\

\section{Helicity flip transitions}

\bf 1. Pure massive Majorana neutrino. 
\rm For massive neutrinos, the helicity states and chiral states are mismatched. The states of ``wrong'' helicity are produced/annihilated by the left-handed weak interactions with amplitudes proportional to $(m_{\nu}/E_{\nu})$ (see Appendix A). Neutrinos with pure Majorana masses have two chiral components, $\nu_{L}$ and $\nu_{R}^{c}$, and the ``wrong'' helicity states are the ordinary right-handed anti-neutrinos.\footnote{For Dirac neutrinos, the ``wrong'' helicity states are the sterile ($SU(2)$ singlet) ``right handed neutrinos'' $\nu_{R}$.} Therefore, from $\pi^{+}$ decay, $\nu_{\mu R}^{c}$ is produced with a fraction
\begin{equation}
f_{\nu} 
= \frac{m_{\nu}^{2}}{m_{\mu}^{2}} \left[ \frac{m_{\mu}^{2} E_{\pi} (E_{\pi}-p_{\pi})}{(m_{\pi}^{2}-m_{\mu}^{2})^{2}}+\frac{m_{\pi}^{2}+m_{\mu}^{2}}{m_{\pi}^{2}-m_{\mu}^{2}} \right] \sim 5 \times 10^{-10}\left( \frac{m_{\nu}}{1~KeV}\right) ^{2}  
\end{equation}
The numerical value is calculated assuming that $\nu_{\mu}$ in the forward direction has an energy of $1$ $GeV$. When the neutrinos are rescattered by nucleons, two reactions produce $\mu^{+}$ events: 
\begin{equation}
\label{e1}
\nu_{\mu}N \rightarrow \mu^{+}X: ~~~ \sigma^{\nu N\rightarrow \mu^{+} X}
            = \frac{G_{F}^{2}}{24 \pi}\frac{m_{\nu}^{2}}{E_{\nu}^{2}}
              s \frac{E_{\nu}}{M} (U^{p}+D^{p})
\end{equation}
is the ``wrong'' helicity interaction, and 
\begin{equation}
\label{e2}
\bar{\nu}_{\mu}N \rightarrow \mu^{+}X: ~~~ \sigma^{\bar{\nu} N\rightarrow \mu^{+} X}
            = \frac{G_{F}^{2}}{2 \pi}s \left( \frac{U^{p}+D^{p}}{3}+2\bar{Q}^{p} \right)    
\end{equation}
is the right helicity interaction, where $s=2ME_{\nu}$, and 
\begin{equation}
 U^{p}\equiv \int^{1}_{0}xu^{p}(x)dx;~~
 D^{p}\equiv \int^{1}_{0}xd^{p}(x)dx;~~
 \bar{Q}^{p}\equiv \frac{\bar{U}^{p}+\bar{D}^{p}}{2} \equiv \int^{1}_{0}x\bar{q}^{p}(x)dx
\end{equation}
are the fractions of momentum carried by $u$, $d$ and anti-quarks in the proton, respectively. (The cross sections are per nucleon for an isoscalar target.) Therefore, the total cross section for $\mu^{+}$ events is
\begin{equation}
\sigma^{\mu^{+}} = f_{\nu} \sigma^{\bar{\nu_{\mu}} N\rightarrow \mu^{+} X} + \sigma^{\nu_{\mu} N\rightarrow \mu^{+} X}.
\end{equation}
Similarly, the cross section for $\mu^{-}$ events is 
\begin{equation}
\label{e3}
\nu N \rightarrow \mu^{-} X: ~~~\sigma^{\mu^{-}}
            = \frac{G_{F}^{2}}{2 \pi}s(U^{p}+D^{p}+\frac{2}{3}\bar{Q}^{p})
\end{equation}

Using $\epsilon \equiv \frac{2 \bar{Q}^{p}}{U^{p}+D^{p}} \sim 0.125$ from 
other experiments, for a typical value $E_{\nu} \sim 1~GeV$ the ratio of $\mu^{+}$ events and $\mu^{-}$ events is approximately 
\begin{equation}
R= \frac{\sigma^{\mu^{+}}}{\sigma^{\mu^{-}}} = 2\times10^{-10}\left( \frac{m_{\nu}}{1~KeV} \right) ^{2}
\end{equation}

For low energy $\nu_{e}$ ($\sim 5~MeV$), we don't assume a particular source, because the low energy neutrinos could come from various reactions, e.g., muon decay from a stopped pion or reactor anti-neutrinos, with the obvious interchange of $\nu_{e} \rightarrow \bar{\nu}_{e}$, $e^{\pm} \rightarrow e^{\mp}$ in the latter case. One can produce the ``wrong'' helicity neutrinos from the original decays. To a good approximation, the rate is $\frac{m_{\nu_{e}}^{2}}{4E_{\nu_{e}}^{2}}$ times the rate to produce the ``right'' helicity neutrinos. Similarly, when $\nu_{e}$ is rescattered from a nucleon target, the ``wrong'' helicity reaction can again take place, producing $e^{+}$ events, with the same factor $\frac{m_{\nu_{e}}^{2}}{4 E_{\nu_{e}}^{2}}$. Hence, the ratio between $e^{+}$ and $e^{-}$ events after rescattering is
\begin{equation}
R \sim \frac{m_{\nu_{e}}^{2}}{2 E_{\nu_{e}}^{2}}\frac{\sigma^{\bar{\nu_{e}}N}}{\sigma^{\nu_{e} N}} \sim 10^{-14},
\end{equation}
assuming that $m_{\nu_{e}}\sim 1~ eV$, $E_{\nu_{e}}\sim 5~ MeV$ and $\sigma^{\bar{\nu_{e}}N}/\sigma^{\nu_{e} N} \sim 1$.  \\

\bf 2. Spin precession in a transverse $B$ field in matter. 
\rm In a transverse magnetic field, the nonzero transition moments of a Majorana neutrino can induce $\nu_{i L} \rightarrow \nu^{c}_{j R}$ transitions between families. Assuming $\nu_{e}$ and $\nu_{\mu}$ as the two families, the evolution equations for the four fields are \cite{s6}
\begin{equation}
i\frac{d}{dt}
\left[\begin{array}{c}
\nu_{e L} \\ \nu_{\mu L} \\ \nu_{e R}^{c} \\ \nu_{\mu R}^{c} 
\end{array}\right]= 
\left[\begin{array}{cccc}
0 & \frac{\Delta m_{21}^{2}}{4E_{\nu}}\sin{2 \theta} & 0 & \mu_{\nu}^{*}B \\ 
\frac{\Delta m_{21}^{2}}{4E_{\nu}}\sin{2 \theta} & \frac{\Delta m_{21}^{2}}{2E_{\nu}}\cos{2 \theta} & -\mu_{\nu}^{*}B & 0 \\
0 & -\mu_{\nu}B & 0 & \frac{\Delta m_{21}^{2}}{4E_{\nu}}\sin{2 \theta} \\
\mu_{\nu}B & 0 & \frac{\Delta m_{21}^{2}}{4E_{\nu}}\sin{2 \theta} & \frac{\Delta m_{21}^{2}}{2E_{\nu}}\cos{2 \theta} 
\end{array}\right]
\left[ \begin{array}{c}
\nu_{e L} \\ \nu_{\mu L} \\ \nu_{e R}^{c} \\ \nu_{\mu R}^{c} 
\end{array}\right]
\end{equation}
where $\theta$ is the vacuum mixing angle, $\Delta m_{21}^2$ is the mass square difference between the two neutrino mass eigenstates, and $E_{\nu}$ is the neutrino energy. The equation can be solved exactly. The transition probability from $\nu_{\mu L}$ to $\nu_{e R}^{c}$ is: 
\begin{equation}
P(\nu_{\mu L} \rightarrow \nu_{e R}^{c})= P_{0} \sin^{2}{\left( \sqrt{4|\mu_{\nu}|^{2}B^{2}+\left(\frac{\Delta m^{2}}{2E_{\nu}}\right) ^{2}}\frac{t}{2} \right)} , 
\end{equation} 
where,
\begin{equation}
P_{0}=\frac{4|\mu_{\nu}|^{2}B^{2}}{4|\mu_{\nu}|^{2}B^{2}+(\frac{\Delta m^{2}}{2E_{\nu}})^{2}}.
\end{equation}
We have neglected second order effects which can lead to the transition $\nu_{\mu L} \rightarrow \nu_{\mu R}^{c}$. 

Suppose that $|\mu_{\nu}|$ is at its experimental limit and that in an experiment there is a $10 T$ transverse $B$ field ($|\mu_{\nu} B| \sim 4.4 \times 10^{-13}~eV$) between the neutrino source and the target. If $B$, $\Delta m^{2}$ and $L$ are sufficiently small so that $\left( \sqrt{4|\mu_{\nu}|^{2}B^{2}+\left(\frac{\Delta m^{2}}{2E_{\nu}}\right) ^{2}}\frac{L}{2} \right) \ll 1$, the transition probability is simply $|\mu_{\nu}BL|^{2}$, independent of $\Delta m^{2}$ and $E_{\nu}$. For example, if we assume $\Delta m^{2} \sim 10^{-5}~eV^{2}$ from the solar neutrino solution \cite{s11}, and $E_{\nu} \sim 1~GeV$, the probability of $\nu_{\mu L}$ converting into $\nu_{e R}^{c}$ at $L \sim 1~km$ is $\sim |\mu_{\nu}BL|^{2}\sim 5 \times 10^{-6}$. If $\left( \sqrt{4|\mu_{\nu}|^{2}B^{2}+\left(\frac{\Delta m^{2}}{2E_{\nu}}\right) ^{2}}\frac{L}{2} \right) \sim 1$, $P_{0}$ is of order $1$ when $\Delta m^{2}/2 E_{\nu}<2|\mu_{\nu}|B$; otherwise , $P_{0}$ is suppressed. Hence, in most of the neutrino experiments a broad spectrum of $E_{\nu}$ can contribute to the conversion rate with fixed $\Delta m^{2}$. 

After the neutrinos are rescattered by the nuclear target, the ratio between the $e^{+}$ and $\mu^{-}$ events is: 
\begin{equation}
\frac{e^{+} events}{\mu^{-} events} = P(\nu_{\mu L}\rightarrow \nu_{e R}^{c}) \frac{\sigma^{\bar{\nu_{e}}N}}{\sigma^{\nu_{\mu} N}}= P\frac{\frac{U^{p}+D^{p}}{3}+2\bar{Q}^{p}}{U^{p}+D^{p}+\frac{2}{3}\bar{Q}^{p}} \approx 0.44 P 
\end{equation}
For $\Delta m^{2}\sim 10^{-5}$, the ratio is about $2\times10^{-6}$.

When $\left( \sqrt{4|\mu_{\nu}|^{2}B^{2}+\left(\frac{\Delta m^{2}}{2E_{\nu}}\right) ^{2}}\frac{L}{2} \right) \sim 1$ and $\Delta m^{2}/2 E_{\nu}>2|\mu_{\nu}|B$, the transition probability can be strongly suppressed by the non-degenerate masses. There are two possible solutions to this. One is to use a neutrino beam with higher energy to test larger $\Delta m^{2}$. Another lies in the fact that $\nu_{e} / \bar{\nu}_{e}$ and $\nu_{\mu} / \bar{\nu}_{\mu}$ have different interactions with matter \cite{s12}, which might be able to compensate for the mass difference at a specific resonance matter density. The transition probability to $\nu_{eR}^{c}$ at time $t$ in matter is (for simplicity, we assume zero mixing angle between neutrino families.),
\begin{equation}
P(\nu_{\mu L} \rightarrow \nu_{e R}^{c}) =
P_{0}^{'}  \sin^{2}{\left( \sqrt{4|\mu_{\nu}|^{2}B^{2}+\left( \frac{\Delta m^{2}}{2E_{\nu}}+a_{\nu_{\mu}}+a_{\nu_{e}} \right) ^{2}}\frac{t}{2} \right)},
\end{equation}
where the different matter potentials are given (for a neutral unpolarized medium) by
\begin{equation}
a_{\nu_{e}}= \frac{G_{F}}{\sqrt{2}}(2N_{e}-N_{n}), ~a_{\nu_{\mu}}= \frac{G_{F}}{\sqrt{2}}(-N_{n}),
\end{equation}
in which $N_{e}$ and $N_{n}$ are respectively the electron and neutron number densities, and
\begin{equation}
P_{0}^{'}=  \frac{4|\mu_{\nu}|^{2}B^{2}}{4|\mu_{\nu}|^{2}B^{2}+(\frac{\Delta m^{2}}{2E_{\nu}}+a_{\nu_{\mu}}+a_{\nu_{e}})^{2}}. 
\end{equation}
A cancellation will occur when 
\begin{equation}
\frac{\Delta m^{2}}{2E_{\nu}}=-a_{\nu_{\mu}}-a_{\nu_{e}}=2(1-\frac{N_{e}}{N_{n}})\frac{G_{F}}{\sqrt{2}}N_{n},  
\end{equation} 
and the suppressed transition rate will be increased. For example, $^{56}_{26}Fe$ at room temperature has $N_{n}=2.54\times10^{24}/cm^3$, and $2(1-\frac{N_{e}}{N_{n}})\approx 0.3$, yielding $-a_{\nu_{e}}-a_{\nu_{\mu}} \sim 6\times 10^{-14}~eV$.   

For low energy $\nu_{e}$ neutrinos\footnote{The $\nu^{c}_{eR} \rightarrow \nu_{\mu L}$ transition probability is the same as that for $\nu_{\mu L} \rightarrow \nu^{c}_{eR}$.}, the probability of $\nu_{e L} \rightarrow \nu_{\mu R}^{c}$ is the same as $P(\nu_{\mu L} \rightarrow \nu_{e R}^{c})$ in transverse $B$ field in vacuum. Here we consider an intermediate neutrino energy\footnote{ Or, $\nu_{e L}$ can oscillate into $\nu_{e R}^{c}$ through second order effects, with probability $\propto |\mu_{\nu}B|^{2}\sin^{2}{2 \theta_{e}}.$} $E_{\nu} \sim 150~ MeV$ so that the $\nu_{\mu R}^{c}$ produced from oscillations can be rescattered into $\mu ^{+}$. Therefore, if $\Delta m^{2} \sim 10^{-5}~eV^{2}$, the probability of having a $\nu_{\mu R}^{c}$ at distance $1~km$ is about $5 \times 10^{-6}$.

After rescattering, the ratio between the $\mu^{+}$ events and $e^{-}$ events is: 
\begin{equation}
\frac{\mu^{+} events}{e^{-} events} = P(\nu_{e L}\rightarrow \nu_{\mu R}^{c}) \frac{\sigma^{\bar{\nu}_{\mu}N}}{\sigma^{\nu_{e} N}} 
\end{equation}
For $\Delta m^{2}\sim 10^{-5}$ and $\sigma^{\bar{\nu}_{\mu}N}/\sigma^{\nu_{e} N} \sim 1$, the ratio is about $5\times 10^{-6}$. \\

\bf 3. Neutrino decay. 
\rm In the model proposed by Chikashige, Mohapatra and Peccei \cite{s7}, an additional higgs singlet field $\phi$ is introduced to couple the $SU(2)$ singlet $\nu_{L}^{c}$ to its conjugate $\nu_{R}$ with a Yukawa coupling constant $h_{2}$. The nonzero vacuum expectation value of $\phi$ spontaneously breaks the global lepton number and generates a Majorana mass term: 
\begin{equation}
L_{mass}= -(\bar{\nu}_{L}~\bar{\nu}^{c}_{L})\left(
\begin{array}{cc}
0 & m \\
m & M
\end{array}\right)\left(
\begin{array}{c}
\nu^{c}_{R} \\
\nu_{R}
\end{array}\right) -h.c.~,
\end{equation}
where $m$ is the Dirac mass and $M=h_{2}\langle \phi \rangle$ is the Majorana mass. It also results in a massless Goldstone boson, the Majoron $\chi$, and interactions between the Majoron and neutrinos. We assume that $m$ and $M$ are comparable (the usual seasaw model makes the opposite assumption, $M\gg m$ \cite{s13}). The two mass eigenstates $\nu_{1}$ and $\nu_{2}$ are
\begin{equation}
\left(
\begin{array}{c}
\nu_{1L} \\
\nu^{c}_{2L}
\end{array}\right) = \left(
\begin{array}{cc}
\cos{\theta} & -\sin{\theta}\\
\sin{\theta} & \cos{\theta}
\end{array}\right) \left(
\begin{array}{c}
\nu_{L}\\ \nu^{c}_{L}
\end{array} \right); ~~
\left(
\begin{array}{c}
\nu^{c}_{1R} \\
\nu_{2R}
\end{array}\right) = \left(
\begin{array}{cc}
\cos{\theta} & -\sin{\theta}\\
\sin{\theta} & \cos{\theta}
\end{array}\right) \left(
\begin{array}{c}
\nu^{c}_{R}\\ \nu_{R}
\end{array} \right)
\end{equation}
where $\tan{\theta}=\frac{2m}{M+\sqrt{M^2+4m^2}}$. The masses of the two eigenstates are comparable. The rate for the decay of the heavier neutrino into the lighter one and the Majoron is \cite{s7}, 
\begin{equation}
\Gamma(\nu_{2 L}^{c} \rightarrow \nu_{1 R}^{c}\chi)= \frac{h_{2}^{2}}{32 \pi} \sin^{2}{\theta} \cos^{2}{\theta} \frac{m_{\nu_{2}}^{2}}{E_{\nu_{2}}}
\end{equation}
where $E_{\nu_{2}}$ is the neutrino energy.

At time zero, a left-handed neutrino $\nu_{\mu L}$ is produced from a weak decay. $\nu_{\mu L}$ is the superposition of two mass eigenstates:
\begin{equation}
\nu_{\mu L}= \cos{\theta_{\mu}}\nu_{1 L}+ \sin{\theta_{\mu}}\nu^{c}_{2 L}.
\end{equation}
While propagating, the $\nu^{c}_{2L}$ component can decay into $\nu^{c}_{1R}$ and a Majoron with rate $\Gamma_{\mu}$. When the decay product is measured through some weak process, only the weak eigenstate component $\nu^{c}_{\mu R}$ of $\nu^{c}_{1R}$ is relevant since $\nu_{\mu R}$ is sterile. This has probability $\cos^{2}{\theta_{\mu}}$, so that at time $t$ the probability of detecting a $\nu^{c}_{\mu R}$ is 
\begin{equation}
P(\nu_{\mu L} \rightarrow \nu^{c}_{\mu R})= \sin^{2}{\theta_{\mu}}\cos^{2}{\theta_{\mu}}\Gamma_{\mu} t~, 
\end{equation}
assuming $\Gamma_{\mu}t$ is small. 

To constrain the Yukawa coupling constant $h_{2}$, we consider the effect of neutrino decay on the MSW resonance condition for the analogous $\nu_{eL} \rightarrow \nu_{e L}^{c}$ case. For definiteness, we assume that an MSW resonance occurs for $\nu_{eL}\rightarrow \nu_{eL}^{c}$ in the sun with $\Delta m^{2} \sim 10^{-5}~ eV^{2}$ and $ \sin^{2}{2\theta_{e}} \sim 0.01$ \cite{s11}. With matter, the evolution equation is
\begin{equation}
i\frac{d}{dt} \left(
\begin{array}{c}
\nu_{e L} \\ \nu^{c}_{e L}
\end{array} \right) = \left(
\begin{array}{cc}
a_{\nu_{e}} & (\frac{\Delta m^{2}}{4 E_{\nu}}-i\frac{\Gamma_{e}}{4})\sin{2\theta_{e}} \\
(\frac{\Delta m^{2}}{4 E_{\nu}}-i\frac{\Gamma_{e}}{4})\sin{2\theta_{e}} & (\frac{\Delta m^{2}}{2 E_{\nu}}-i\frac{\Gamma_{e}}{2})\cos{2\theta_{e}} 
\end{array}\right) \left(
\begin{array}{c}
\nu_{e L} \\ \nu^{c}_{e L}
\end{array} \right).
\end{equation}
(Of course, $\Delta m^{2}$, $\theta_{e}$, $\Gamma_{e}$ and $h_{2e}$ need not be the same as for the $\nu_{\mu L}$ case.) At resonance, $a_{\nu_{L}}= \frac{\Delta m^{2}}{2 E_{\nu}}\cos{2 \theta_{e}}>0$, and significant transitions can occur provided 
\begin{equation}
|\frac{\Gamma_{e}}{2}\cos{2\theta_{e}}|^{2}\ll |2(\frac{\Delta m^{2}}{4 E_{\nu}}-i\frac{\Gamma_{e}}{4})\sin{2\theta_{e}}|^{2}.
\end{equation}
Therefore, 
\begin{equation}
\Gamma_{e} \ll \frac{\Delta m^{2}}{E_{\nu}}\sin{2\theta_{e}}.
\end{equation}
For $\Delta m^{2} \sim 10^{-5}~ eV^{2}$ and $ \sin^{2}{2\theta_{e}} \sim 0.01$ from the MSW solar neutrino solution, and typical $E_{\nu_{e}} \sim 1~MeV$, $m_{\nu_{e}} \sim 1~eV$, the constraint on $\Gamma_{e}$ gives $h_{2e}^{2} \ll 0.1$. Of course, $m_{\nu_{e}} \sim 1~eV$ is a very optimistic assumption - the bound on $h_{2e}$ will be looser for smaller $m_{\nu_{e}}$. 

The limit on the mixing angle for $\nu_{\mu}$ given by disappearance experiments is $\sin^{2}{2 \theta_{\mu}}<0.02$ for $\Delta m^{2}= 100~ eV^{2}$. We assume that the limit $h_{2e}^{2}\ll 0.1$ for electron neutrinos applies also to muon neutrinos, i.e., that the higgs singlet field couples to neutrinos from different families with similar strength, and take the typical distance between the neutrino source and the target in the laboratory environment to be $1~km$. After the neutrinos are rescattered from nucleons, the ratio between $\mu^{+}$ and $\mu^{-}$ events,
\begin{equation}
R= P(\nu_{\mu L}\rightarrow \nu_{\mu R}^{c}) \frac{\sigma^{\bar{\nu_{\mu}}N}}{\sigma^{\nu_{\mu} N}}\approx 0.44 P,
\end{equation} 
is less than $4\times 10^{-7}$ for the assumed parameter values. 

For low energy $\nu_{e}$, the transition probability at time t is
\begin{equation}
P(\nu_{e L} \rightarrow \nu^{c}_{e R})= \sin^{2}{\theta_{e}}\cos^{2}{\theta_{e}}\Gamma_{e} t
\end{equation}
where $\Gamma_{e}=\frac{h_{2}^{2}}{32 \pi} \sin^{2}{\theta_{e}} \cos^{2}{\theta_{e}} \frac{m_{\nu_{2}^{2}}}{E_{\nu_{2}}}$, analogous to the $\nu_{\mu}$ case with the mixing angle $\theta_{e}$ for $\nu_{e}$. If $\sin^{2}{2\theta_{e}} \sim 0.01$ for $\Delta m^{2}\sim 10^{-5} eV^{2}$, $m_{\nu_{e}} \sim 1~ eV$ and $E_{\nu_{e}}\sim 5~ MeV$, the $\nu_{e L}$ to $\nu_{e R}^{c}$ transition probability after the neutrino travels $1~ km$ is less than $10^{-4}$. Therefore, the ratio between $e^{+}$ events and $e^{-}$ events after rescattering is less than $P r \sim 10^{-4}$. \\      

\section{Helicity non-flip models}

\bf 1. $SU(2)_{R}\times SU(2)_{L} \times U(1)$ model. 
\rm As discussed in the Introduction, for neutrinos with comparable Dirac and Majorana masses, $2^{nd}$ class oscillations can occur. They can be quite efficient for converting neutrinos into anti-neutrinos. However, they create $SU(2)$ singlet states which interact only via mixing or new interactions. In the $SU(2)_{R}\times SU(2)_{L} \times U(1)$ model, $\nu^{c}_{L}$ and $\mu^{+}_{L}$ are grouped into an $SU(2)_{R}$ doublet, and the right-handed interactions are carried by gauge bosons $W_{R}^{\pm}$ \cite{s8}. It is also possible that $W_{L}$ and $W_{R}$ and their mass eigenstates are mismatched, which causes a mixture of right-handed and left-handed interactions. The complete 4-Fermi interaction for the charged weak current of the  
$SU(2)_{L}\times SU(2)_{R}\times U(1)$ model is: \cite{s9} 
\begin{equation}
H= \frac{4\hat{G_{F}}}{\sqrt{2}}(a J^{\dagger}_{L\rho} J^{\rho}_{L} + b 
J^{\dagger}_{L\rho} J^{\rho}_{R} + c J^{\dagger}_{R\rho} J^{\rho}_{L}+ 
d J^{\dagger}_{R\rho} J^{\rho}_{R}) 
\end{equation}
where, 
\begin{eqnarray*}
\frac{\hat{G_{F}}}{\sqrt{2}} & = & \frac{g_{L}^{2}\cos^{2}{\xi}}{8M_{1}^{2}}; 
\\
J^{\dagger}_{L,R\rho} & = & \bar{u}_{L,R}\gamma_{\rho}U^{L,R}d_{L,R}+\bar{\nu}_{L,R}
\gamma_{\rho}V^{L,R}e_{L,R};  \\
a & = & 1+ \beta\tan^{2}{\xi}; \\
b^{*} & = & c= e^{i\omega}(g_{R}/g_{L})\tan{\xi}(1-\beta); \\
d & = & (g_{R}^{2}/g_{L}^{2})(\tan^{2}{\xi}+\beta);
\end{eqnarray*}
with $\beta\equiv M_{1}^{2}/M_{2}^{2}$, where $M_{1}$ and $M_{2}$ are $W$ boson masses. $U^{L}$ is the CKM matrix for quarks, and $U^{R}$ is its analog for $J_{R}$. We ignore neutrino mass effects in the rescattering, so the lepton mixing matrix $V^{L}$ can be taken to be the identity. 

Therefore, $\nu_{\mu L}^{c}$ can be scattered into $\mu^{+}$ through right-handed interactions. There are two reactions to produce $\mu^{+}$.  
$\nu^{c}_{\mu L}N\rightarrow \mu^{+}X$ through right-handed currents, 
 \begin{equation}
\sigma^{\nu^{c}_{\mu L}N(RR)}= \frac{G_{F}^{2}}{2 \pi}|d|^{2} 
|U^{R}_{du}|^{2} s (\frac{U^{p}+D^{p}}{3}+2\bar{Q}^{p}), 
\end{equation}
and through the mixing of right-handed and left-handed currents: 
\begin{equation}
\sigma^{\nu^{c}_{\mu L}N(RL)}= \frac{\hat{G}_{F}^{2}}{2 \pi}|c|^{2} |U^{L}_{du}|^{2} s (U^{p}+D^{p}+\frac{2}{3}\bar{Q}^{p}).
\end{equation}
For $|b|$, $|d|$ $\ll a$, $\mu^{-}$ production $\nu_{\mu}N \rightarrow \mu^{-}X$ proceeds through ordinary left-handed currents:
\begin{equation}
\sigma^{\nu_{\mu L} N(LL)}
            = \frac{\hat{G}_{F}^{2}}{2 \pi}|a|^{2}
|U^{L}_{du}|^{2}s(U^{p}+D^{p}+\frac{2}{3}\bar{Q}^{p}). 
\end{equation}  

Therefore the ratio between $\mu^{+}$ events and $\mu^{-}$ events is: 
\begin{equation}
R= \frac{ P(\nu_{\mu L} \rightarrow 
\nu^{c}_{\mu L}) [ \sigma^{\nu^{c}_{\mu L}N(RR)} + \sigma^{\nu^{c}_{\mu L}N(RL)} ]}
{P(\nu_{\mu L} \rightarrow \nu_{\mu L}) \sigma^{\nu_{\mu L}N(LL)} }
\end{equation}   
We work in the manifest (or pseudo-manifest) left-right symmetric model, in which $g_{L}=g_{R}$ and $|U^{R}_{ud}|=|U^{L}_{ud}|$, and use the limits \cite{s9} for the model parameters, $|\xi| < 0.003$, $\beta<0.004$. Assuming 
 $\sin^{2}{2\theta_{\mu}}<0.02$ for $\Delta m^{2}=100~eV^{2}$ from disappearance experiments, one finds 
\begin{equation}
R \sim 3\times 10^{-7} \sin^{2}{(\Delta/2)},
\end{equation}
which is less than $ 3\times 10^{-7}$ for $L\sim 1~km$ and $E_{\nu} \sim 1~ GeV$. 

In the low energy $\nu_{e}$ case, when $\nu_{e L}^{c}$ is rescattered from nucleons following $2^{nd}$ class oscillations, there are again two reactions to produce $e^{+}$. $\nu^{c}_{e L}N\rightarrow e^{+}N^{'}$ through right-handed currents and the mixing between right-handed and left-handed currents, with cross sections $\sigma^{\nu^{c}_{e L}N(RR)}$ and $\sigma^{\nu^{c}_{e L}N(RL)}$ respectively. Full estimates require detailed study of the nuclear transitions and are beyond the scope of this paper. However, to obtain rough estimates, it is useful to define reduced cross sections with the suppression factors $|c|^{2}$ and $|d|^{2}$ removed:

\begin{equation}
\sigma^{\nu^{c}_{e L}N(RR)}= |d|^{2} \hat{\sigma}^{\nu_{eL}^{c}N(RR)}.
\end{equation}
Analogously, 
\begin{equation}
\sigma^{\nu^{c}_{e L}N(RL)}= |c|^{2} \hat{\sigma}^{\nu_{eL}^{c}N(RL)}.
\end{equation}
For $e^{-}$ events, $\nu_{e}N \rightarrow e^{-}X$ occurs through left-handed currents:
 \begin{equation}
\sigma^{\nu_{e L} N(LL)}= |a|^{2}\sigma^{\nu_{e} N}.
 \end{equation}  
Therefore, the ratio between $e^{+}$ and $e^{-}$ events after rescattering is 
\begin{equation}
R= \frac{ P(\nu_{e L} \rightarrow 
\nu^{c}_{e L}) [ |d|^{2} \hat{\sigma}^{\nu_{eL}^{c}N(RR)}  +|c|^{2} \hat{\sigma}^{\nu_{eL}^{c} N(RL)}  ]}
{P(\nu_{e L} \rightarrow \nu_{e L}) |a|^{2}\sigma^{\nu_{e} N}}
\end{equation}   
Putting in limits on the $SU(2)_{L}\times SU(2)_{R} \times U(1)$ parameters, $\sin^{2}{2\theta_{e}} \sim 0.01$ and assuming $\hat{\sigma}^{\nu_{eL}^{c}N(RR)} \sim \hat{\sigma}^{\nu_{eL}^{c}N(RL)} \sim \sigma^{\nu_{e} N}$, the ratio is \begin{equation}
R \sim 10^{-7} \sin^{2}{(\Delta/2)}
\end{equation}
For $\Delta m^{2} \sim 10^{-5}~eV^{2}$, $E_{\nu_{e}}\sim 5~MeV$, and $L\sim 1~km$, R is around $10^{-11}$.  
\\

\bf 2. Models with mixing between ordinary and exotic fermions. 
\rm Many models predict the existence of new fermions with exotic $SU(2)\times U(1)$ assignments. The mass eigenstates associates with them are usually heavy, and the mixings between the ordinary light states and the heavy exotic states are small \cite{s10}. In a particular model, the Lagrangian mass term is: 
\begin{equation}
L_{mass}= - \left( \bar{\nu}_{L}, \bar{N}_{L}, \bar{N}_{L}^{c}, \bar{\nu}_{L}^{c} \right) M \left( \begin{array}{c}
\nu_{R}^{c} \\ 
N_{R}^{c} \\
N_{R} \\
\nu_{R}
\end{array} \right ) + h.c. 
\end{equation}
where $(\nu_{L},~\nu_{R}^{c})$ is an ordinary doublet neutrino, $(\nu_{L}^{c},~\nu_{R})$ is a singlet, and $(N_{L},~N_{R}^{c})$ and $(N_{L}^{c},~N_{R})$ are new states which may be $SU(2)$ singlets or doublets; $M$ is a $4\times 4$ complex symmetric mass matrix. Diagonalizing $M$, we can find the mass eigenstates and eigenvalues. We assume that at most two of the four eigenstates $\nu_{1}$ and $\nu_{2}$ are light. The weak states are superpositions of mass eigenstates: 
\begin{equation}
\nu_{L}= U_{11}^{'}\nu_{1L}+U_{12}^{'}\nu_{2L}+U_{13}^{'}\nu_{3L}+U_{14}^{'}\nu_{4L}; 
\end{equation}
and the light mass eigenstates are combinations of the four weak eigenstates: 
\begin{equation}
\left\{ \begin{array}{cc}
\nu_{1L}& = U_{L11}\nu_{L}+U_{L12}N_{L}+U_{L13}N_{L}^{c}+U_{L14}\nu_{L}^{c}; \\
\nu_{2L}& = U_{L21}\nu_{L}+U_{L22}N_{L}+U_{L23}N_{L}^{c}+U_{L24}\nu_{L}^{c};  
\end{array}
\right.
\end{equation}
where $U_{L}=U^{'-1}$ is a unitary matrix. From pion decay, only the light components of $\nu_{\mu L}$ can be produced, resulting an incomplete state $\nu_{\mu L}^{'}$. If there are two light states, 
\begin{equation}  
\nu_{\mu L}^{'}= U_{11}^{'}\nu_{1L}+U_{12}^{'}\nu_{2L};
\end{equation}
which is in general not orthogonal to $N_{L}$, $N_{L}^{c}$, or $\nu_{L}^{c}$ \cite{s10A}. The decay rate is $\Gamma_{0}(|U_{11}^{'}|^{2}+|U_{12}^{'}|^{2})$, where $\Gamma_{0}$ is the decay rate without mixing. The state evolves to $\nu_{\mu L}^{'}(t)$ when it reaches the target: 
\begin{equation}
\nu_{\mu L}^{'}=e^{-ipt}(e^{-i\frac{m_{1}^{2}}{2p}t} U_{11}^{'}\nu_{1}+e^{-i\frac{m_{2}^{2}}{2p}t}U_{12}^{'}\nu_{2}).
\end{equation}  
When it is rescattered from a nucleon target, three weak transitions may occur, due to the nonorthogonality of the state, oscillations (for $m_{1}^{2}\neq m_{2}^{2}$), or both:   
\begin{equation}
(\nu_{L} \rightarrow \mu_{L}^{-}); ~ (N_{L} \rightarrow E_{L}^{-}); ~ (N_{L}^{c} \rightarrow E_{L}^{+}) 
\end{equation}
The second and third can occur only if $(N_{L}, ~E_{L}^{-})$ and/or $(N_{L}^{c}, ~E_{L}^{+})$ belong to $SU(2)$ doublets, respectively. Assume that the mixing angle between $E_{L}^{-}$ and $\mu^{-}_{L}$ is $\theta_{\mu L}$; and that between $E_{L}^{+}$ and $\mu^{+}_{L}$ is $\theta_{\mu R}$. The probability of producing a $\mu^{-}$ from rescattering is: 
\begin{equation}
\begin{array}{ccc}
P(\mu^{-}) & = & \left[ |U_{11}^{'}U_{L11}+U_{12}^{'}U_{L21} 
             e^{-i\frac{\Delta m^{2}}{2p}t}|^{2} \cos^{2}{\theta_{\mu L}}
                 \right .  \\ 
           &   & \left.  + |U_{11}^{'}U_{L12}+U_{12}^{'}U_{L22} 
              e^{-i\frac{\Delta m^{2} }{2p}t}|^{2} \sin^{2}{\theta_{\mu L}} 
              \right] \sigma^{\nu N},
\end{array}
\end{equation}
and the probability of producing a $\mu^{+}$ is:
\begin{equation}
P(\mu^{+}) = |U_{11}^{'}U_{L13}+U_{12}^{'}U_{L23} e^{-i\frac{\Delta m^{2} }{2p}t}|^{2} \sin^{2}{\theta_{\mu R}}  \sigma^{\nu N}, 
\end{equation} 
with $\Delta m^{2}= m_{2}^{2}-m_{1}^{2}$. $P(\mu^{+})$ can be non-zero even if $\Delta m^{2}=0$ due to the nonorthogonality of $\nu_{\mu L}^{'}$ and $N^{c}_{L}$. 

In a mirror model, one right-handed doublet and two left-handed singlets are introduced (plus their CP conjugates),  
\begin{equation}
N_{L}, E^{-}_{L}, \left( \begin{array}{c}
N \\
E^{-}
\end{array}\right)_{R}
\end{equation}
The mass matrix is: 
\begin{equation}
\left( \begin{array}{cccc}
0     & d_{1} & s_{1} & d_{2} \\
d_{1} & s_{3} & d_{3} & s_{2} \\
s_{1} & d_{3} & 0     & d_{4} \\
d_{2} & s_{2} & d_{4} & s_{4} 
\end{array} \right ), 
\end{equation}
where $s_{i}$ and $d_{i}$ can be generated by Higgs singlets and doublets, respectively. In particular, $s_{4}$ is a Majorana mass term for $\nu_{R}$ and $d_{2}$ is its Dirac mass term. For definiteness, we assume the seesaw model so that $d_{2}$, $s_{1}, d_{1} \ll s_{4}$ are small mixing terms. We also assume that $N$ has a large Dirac mass, $d_{3} \gg s_{3}$. Finally, we assume that the mixing terms $s_{2}, d_{4}$ are small. Thus, $s_{4}, d_{3}$ are large and all others are perturbations. In this case, there are 3 heavy states and 1 light state with, $m_{1}\sim (2d_{1}s_{1}/d_{3})+(d_{2}^{2}/s_{4}), m_{2} \sim m_{3} \sim d_{3}, m_{4} \sim s_{4}$. The light mass eigenstate, to first order in the perturbations, is: 
\begin{equation}
\nu_{1L}= \nu_{L}-(s_{1}/d_{3}) N_{L} - (d_{1}/d_{3}) N_{L}^{c} -(d_{2}/s_{4})\nu^{c}_{L} 
\end{equation}
Therefore, $U_{11} \sim 1; U_{12}\sim -s_{1}/d_{3}$ and $U_{13}\sim -d_{1}/d_{3}$. Since there is only one light state, there are no oscillations, but $\mu^{+}$ can still be produced due to the non-orthogonality of $\nu^{'}_{L}$ and $N_{L}^{c}$. The ratio of $\mu^{+}$ events and $\mu^{-}$ events is $\frac{|U_{13}|^{2}\sin^{2}{\theta_{\mu R}}}{|U_{11}|^{2}\cos^{2}{\theta_{\mu L}}}$, for $N_{L}$ is sterile in the mirror model. The limit \cite{s10} \cite{s10A} is $|U_{13}|^{2}<0.027$.

For charged leptons, the Lagrangian mass term is:
\begin{equation}
L_{mass}=- \left( \bar{\mu}_{L}^{-}, \bar{E}_{L}^{-}\right) M_{E} \left( \begin{array}{c}
\mu_{R}^{-} \\ 
E_{R}^{-}  \end{array} \right ) + h.c. 
\end{equation}
where $M_{E}=\left( \begin{array}{cc}
d_{5} & s_{1} \\
s_{5} & d_{6} 
\end{array} \right) 
$. $d_{6}$, the Dirac mass term for $E^{-}$, could be much larger than the Dirac mass $d_{5}$ for $\mu^{-}$. If we assume that the mixing terms $s_{1}$ and $s_{5}$ are of the same order as $d_{5}$, there are two mass eigenstates $m_{1}^{2} \sim d_{5}^{2}, m_{2}^{2} \sim d_{6}^{2}$, the mixing angle between $\mu^{-}_{L}$ and $E_{L}^{-}$ is $\theta_{\mu L} \sim s_{1}/d_{6}$, and the mixing angle between $\mu^{-}_{R}$ and $E_{R}^{-}$ is $\theta_{\mu R} \sim s_{5}/d_{6}$. Using the experimental limits on heavy fermions \cite{s10} \cite{s10A}, a loose estimate of the two angles is $ \theta_{\mu R},~ \theta_{\mu L} < m_{\mu}/(90 ~GeV) \sim 0.0014 $. Hence the ratio between $\mu^{+}$ and $\mu^{-}$ events is $R < 4\times 10^{-8}$. 

Similarly, for low energy $\nu_{e}$, the ratio between $e^{+}$ and $e^{-}$ events is then 
\begin{equation}
R=\frac{|U_{13}|^{2}\sin^{2}{\theta_{e R}}}{|U_{11}|^{2}\cos^{2}{\theta_{e L}}} \frac{\sigma^{N_{L}^{c}N}}{\sigma^{\nu_{eL}N}},
\end{equation}
for the mirror model, where the parameters are defined for the electron family, and the cross sections are for light leptons (with the mixing angle removed). The limit \cite{s10} \cite{s10A} is $|U_{13}|^{2}<0.047$ for $\nu_{e}$, and $\theta_{e R}, ~\theta_{e L}<6 \times 10^{-6}$. Therefore, the ratio is $R < 2 \times 10^{-12}$.       

We have done a similar study of the vector doublet model, i.e., $(N,~E^{-})_{L}$ and $(N,~E^{-})_{R}$ are both $SU(2)$ doublets, but no second order effects (in amplitude) were found. 
\\

\section{Discussion}

Models beyond the SM predict lepton number violating transitions $\nu_{L}\rightarrow \nu_{R}^{c}$ or $\nu_{L}^{c}$ through different mechanisms. We have considered five scenarios in which neutrino-anti-neutrino transitions in the laboratory are possible, and two kinds of experiments. One is that high energy muon neutrinos from pion decay are deep-inelastically rescattered from an isoscalar target; the lepton number violation transition manifests itself by producing $\mu^{+}$ events from the rescattering. (One could easily generalize the models to consider changing lepton families as well, yielding $e^{+}$ or $\tau^{+}$.) The results from different scenarios are collected in Table 3. The first three scenarios involve helicity flip into an active anti-neutrino, but the helicity flip rate is small. The last two involve larger transition rates (via second class oscillations) without helicity flip, but the resulting singlet neutrinos can only interact by (small) mixings or new interactions.

\begin{table}
\begin{center}
\begin{tabular}{ccc} 
 model  & parameters & $\frac{\mu^{+}events}{\mu^{-} events}$  \\ \hline
 Pure Majorana  & $m_{\nu_{\mu}}$ & $ < 10^{-10}$ \\ \hline
 Spin precession  & $ |\mu_{\nu_{\mu}}|< 7.4\times10^{-10}\mu_{B}$  & $< 2 \times 10^{-6}$ \\
 in $B_{\perp}$  & $\Delta m^{2} \sim 10^{-5}~ eV^{2}$ & $(L \sim 1~km)$ \\ \hline
 Neutrino Decay  & $h_{2}^{2}< 0.1$, $m_{\nu_{\mu}} \sim 10~ eV$, $\sin^{2}{2\theta_{\mu}} < 0.02$   & $< 4\times 10^{-7}$ \\ \hline
 $ SU(2)_{L}\times SU(2)_{R}\times U(1)$  &  $|\xi_{g}| < 0.003$, $\beta_{g}<0.004$  & $<3\times 10^{-7}$  \\ 
  &   $\sin^{2}{2\theta_{\mu}}<0.02$ for $\Delta m^{2}=100~eV^{2}$ & $ (L \sim 1~km) $ \\ \hline
 Exotic fermions  & $|U_{13}^{2}| < 0.027$, $ \theta_{\mu R}, ~\theta_{\mu L}\sim 0.0014 $  & $< 4\times 10^{-8}$ 
\end{tabular} 
\caption{$\mu^{+}$, $\mu^{-}$ events ratio of high energy $\nu_{\mu}$ ($\sim 1~GeV$) $N$ scattering for five neutrino-antineutrino oscillation scenarios. ($e^{+}$, $\mu^{-}$ events ratio for the spin precession scenario.) }
\end{center}
\end{table}  

In the second case, we have considered a low energy electron neutrino beam and showed the magnitude of the effective $\nu \rightarrow \bar{\nu}$ transition with the ratio between $e^{+}$ and $e^{-}$ events after rescattering, which is listed in Table 4. 

\begin{table}
\begin{center}
\begin{tabular}{ccc} 
 model  & parameters & $\frac{e^{+}events}{e^{-} events}$  \\ \hline
 Pure Majorana  & $m_{\nu_{e}}$ & $ < 10^{-14}$ \\ \hline
 Spin precession  & $ |\mu_{\nu_{\mu}}|< 7.4\times10^{-10}\mu_{B}$  & $ < 5 \times 10^{-6}$ \\
 in $B_{\perp}$  & $\Delta m^{2} \sim 10^{-5}~ eV^{2}$ & $(L \sim 1~km)$  \\ \hline
 Neutrino Decay  & $h_{2}^{2}< 0.1$, $m_{\nu_{e}} \sim 1~ eV$, $\sin^{2}{2\theta_{e}} < 0.01$   & $< 10^{-4}$ \\ \hline
 $ SU(2)_{L}\times SU(2)_{R}\times U(1)$  &  $|\xi_{g}| < 0.003$, $\beta_{g}<0.004$  & $< 10^{-11}$  \\ 
  &   $\sin^{2}{2\theta_{e}}<0.01$ for $\Delta m^{2}=10^{-5}~eV^{2}$ & $ (L \sim 1~km) $ \\ \hline
 Exotic fermions  & $|U_{13}^{2}| < 0.047$, $ \theta_{e R},~\theta_{e L}\sim 6\times 10^{-6} $ & $< 2 \times 10^{-12}$ 
\end{tabular} 
\caption{$e^{+}$, $e^{-}$ events ratio for low energy $\nu_{e}$ ($\sim 1 ~MeV$) $N$ scattering for five scenarios. (In spin precession, the result is calculated with $E_{\nu}\sim 150~MeV$.)}
\end{center}
\end{table}

From the table, there can be a non-zero probability for transitions between neutrinos and anti-neutrinos. The upper limit of the transition is $10^{-6}$ $\sim$ $10^{-10}$ for high energy neutrinos ($\sim 1~GeV$) for the various models, depending on model parameters, whose limits are set by other experiments and analysis. The present statistics on high energy $\nu_{\mu} N$ scattering CC events is $\sim 1.7\times 10^{5}$ for CDHS/CHARM, and $\sim 1.1\times 10^{6}$ for CCFR \cite{s14}. Even statistically, these experiments are not sensitive to the upper limit $10^{-6}$ for the $\mu^{+}/\mu^{-}$ event ratio. Furthermore, in these experiments, in which $\pi^{+}$ are produced from proton scattering, $\pi^{-}$ are also produced, with a cross section about one eighth of that of $\pi^{+}$ production. Those $\pi^{-}$, which can decay into $\mu^{-}$ and $\bar{\nu}_{\mu}$, give a significant background for the $\nu_{\mu} \rightarrow \bar{\nu}_{\mu}$ transition effect. Even if tagging or other techniques are used to distinguish $\bar{\nu}_{\mu}$ from $\pi^{-}$, there are decays like $\mu^{+} \rightarrow e^{+} \nu_{e} \bar{\nu}_{\mu}$ in the $\pi^{+}$ beam, which is also significant compared to the transition effects. Considering the flux and the background in the present high energy $\nu_{\mu}$ experiments, it is unlikely that the effects could be observed.  

The upper limit of the transition for the low energy $\nu_{e}$ ($\sim 5~MeV$) is about $10^{-4}\sim 10^{-14}$. We note that the analysis and results are also true for reactor $\bar{\nu}_{e}$ to $\nu_{e}$ transitions. Nuclear power reactors are the most intense sources of $\bar{\nu}$ available on earth. The event rates in such experiments are restricted by the small cross sections for low energy neutrinos and the $1/L^{2}$ decrease of the neutrino flux when the detector is away from the source. For example, BUGEY \cite{s15} with thermal power $2.8~GW$ has event rates $62.62~h^{-1}$ at $L\sim 15~m$, $15.39~h^{-1}$ at $L \sim 40~m$ and $1.38~h^{-1}$ at $L \sim 90~m$; CHOOZ \cite{s16} with $8.5~GW$ has event rate $25~d^{-1}$ at $L=1~km$, etc. Hence, the event rates of present reactor experiments are too low to observe any $\bar{\nu} \rightarrow \nu$ transition effect. Neutrinos from $\pi$ and $\mu$ decay at rest from LAMPF can be used to study $\nu_{e} \rightarrow \bar{\nu}_{e}$ or $\nu_{\mu} \rightarrow \bar{\nu}_{e}$ (in a magnetic field). A loose estimation from present data of LSND \cite{s17} shows that the event rate is again too low, besides the severe problem of background signals.          
Very intense neutrino sources from $\pi$ and $\mu$ decay might become available from a possible future muon collider or dedicated storage ring \cite{s18}. Even if enough events were available to produce a significant number of neutrino-antineutrino transitions, it would be extremely difficult to separate them from standard model backgrounds or from likely first class oscillations; e.g., for $\mu^{+} \rightarrow e^{+} \nu_{e} \bar{\nu}_{\mu}$, a $\bar{\nu}_{e}$ from $\nu_{e} \rightarrow \bar{\nu}_{e}$ would be hard to distinguish from one due to $\bar{\nu}_{\mu} \rightarrow \bar{\nu}_{e}$.     

Thus, while $\nu \rightarrow \bar{\nu}$ transitions are possible in principle, it is unlikely that they can be observed in foreseeable laboratory experiments.

\section{Acknowledgments}

It is a pleasure to thank Boris Kayser for discussions which led to this investigation. This work was supported in part by U.S. Department of Energy Grant No. DOE-EY-76-02-3071. 

\section{Appendix A}
The Dirac equation for massive fermions is: 
\begin{equation}
(i \gamma^{\mu}\partial_{\mu}-m)\psi(x) = 0. 
\end{equation}
In the Weyl representation, 
\begin{equation}
 \gamma^{\mu}=\left(
  \begin{array}{cc}
   0 & \sigma^{\mu} \\
   \bar{\sigma}^{\mu}&0
  \end{array}
 \right), ~  
 \gamma^{5}=\left(
  \begin{array}{cc}
   -1&0\\
   0&1
  \end{array}
 \right),
\end{equation}
in which $\sigma^{\mu}=(1,\sigma^{i})$,~ 
 $\bar{\sigma}^{\mu}=(1,-\sigma^{i})$,~ and $\sigma^{i}$ are Pauli matrices.
The positive and negative frequency solutions are, respectively,
\begin{equation}
 u(p)= \left(
	\begin{array}{c}
	\sqrt{p\cdot\sigma} \xi \\
	\sqrt{p\cdot\overline{\sigma}} \xi
	\end{array}
       \right), ~
 v(p)= \left(
	\begin{array}{c}
	\sqrt{p\cdot\sigma} \eta \\
	-\sqrt{p\cdot\overline{\sigma}} \eta
	\end{array}
       \right),
\end{equation}
where $\xi$ and $\eta$ are 2-component spinors, from which the helicities of the particles are determined.

The weak current of an $e^{-}\rightarrow \nu_{e}$ transition is  
\begin{equation}
J^{\mu}=\bar{u}(p_{\nu})\gamma^{\mu}\frac{1-\gamma_{5}}{2}u(p_{e}). 
\end{equation}
If the initial electron is polarized along the $-z$ direction $\xi^{e}=(0,1)^{T}$, the current for producing a final state neutrino with negative helicity $\xi^{\nu}=(0,1)^{T}$ is    
\begin{equation}
 J^ {\mu} =\left\{
	  \begin{array}{ll}
	   \pm 2\sqrt{E_{e}E_{\nu}}; & \mu=0,3 \\
	   0; & \mu=1,2
	  \end{array}
         \right.   
\end{equation}
and for a positive (``wrong'') helicity neutrino $\xi^{\nu}=(1,0)^{T}$ , is 
\begin{equation}
 J^{\mu}=\left\{
	  \begin{array}{llll}
	   0; & \mu=0,3 \\
           \pm 2\sqrt{E_{e}E_{\nu}}\frac{m_{\nu}}{2E_{\nu}}; & \mu=1,2 
	  \end{array}
         \right.   
\end{equation}
Therefore, the probability for producing the ``wrong'' helicity neutrino from the left-handed weak current is proportional to $(\frac{m_{\nu}}{2E_{\nu}})^{2}$. 

For a Majorana neutrino, the mass term in the Lagrangian is: $-L_{mass}=\frac{1}{2}m_{M} \bar{\psi}_{L}\psi^{c}_{R} + h.c.$ in terms of chiral fermion fields. $\psi_{L}$ and $\psi^{c}_{R}$ can be combined to form a two component Majorana neutrino: $\Psi=\psi_{L}+\psi_{R}^{c}$ so that $-L_{mass}=\frac{1}{2}m_{M}\bar{\Psi}\Psi$, and $\Psi$ satisfies the Dirac equation with mass $m_{M}$: 
\begin{equation}
(i \gamma^{\mu}\partial_{\mu}-m_{M})\Psi(x) = 0, 
\end{equation}
Therefore, the probability for producing the ``wrong'' helicity state from the left-handed weak current for massive Majorana neutrinos is also proportional to $(\frac{m_{\nu}}{2E_{\nu}})^{2}$, as shown in the example, except that the ``wrong'' helicity state produced in the Majorana neutrino case is approximately a right-handed anti-neutrino $\psi^{c}_{R}$, while in the Dirac neutrino case it is a right-handed neutrino $\psi_{R}$.   

\bigskip

\bibliographystyle{prsty}
\bibliography{tdw}

\end{document}